# Studies on $La_{2-x}Pr_xCa_yBa_2Cu_{4+y}O_z$ (x = 0.1 – 0.5, y = 2x) type mixed oxide superconductors

S. Rayaprol[a,*], K. R. Mavani[a], D. S. Rana[a], C. M. Thaker[a], Manglesh Dixit[b], Shovit Bhattacharya[b] and D. G. Kuberkar[a]

[a] *Department of Physics, Saurashtra University, Rajkot – 360 005 (INDIA)*
[b] *Department of Physics, Barkatullah University, Bhopal (INDIA)*

**Abstract**
The $La_{2-x}Pr_xCa_yBa_2Cu_{4+y}O_z$ (LaPrCaBCO) mixed oxides have been studied for their structural and superconducting properties using X-ray diffraction (XRD), d. c. resistivity, d. c. susceptibility and iodometric double titration. All the LaPrCaBCO samples for x = 0.1 – 0.5, exhibit tetragonal crystalline structure with P 4/mmm space group as determined by Rietveld analysis of the X-ray diffraction data. With increasing x, enhancement in $T_c$ is observed, which is quite interesting for Pr substituted high $T_c$ oxides. Maximum $T_c \sim 58$ K has been observed for x = 0.5 (La-2125 stoichiometry). The results of structural studies and superconducting property measurements are presented in light of increase in $T_c$ in LaPrCaBCO system with increasing Pr concentration.

[*] Corresponding Author,  Present Address: DCMP & MS, TIFR, Mumbai, India (sudhindra@rayaprol.com)

1. **INTRODUCTION**

The substitution of Pr in R-123 (R = rare earth) and related systems has been widely studied, for its role in suppressing superconductivity either due to hybridization between Pr (4f) and conducting $CuO_2$ (2p) plane leading to localization of mobile charge carriers and/or hole filling by fluctuating valence of Pr (3+ or 4+) at rare earth or even at Ba (Barium) site [1-5]. Even though many studies have been carried out on these compounds, yet the correct explanation for such a phenomenon is still a matter of debate. Majority of Pr substituted R-123 and derived systems have shown decrease in $T_c$ with increasing Pr concentration. The Pr-substituted R-123 compounds also show rare earth ion effect and for every $x_{cr}$ (critical Pr concentration) with respect to the R ion, superconductivity is totally suppressed [6, 7].

Our studies on the $La_{2-x}R_xCa_yBa_2Cu_{4+y}O_z$ (La-2125) type of R-123 derived compounds have shown enhancement in $T_c$ with increasing x, with maximum $T_c \sim 78$ K achieved with different R, for x = 0.5 compound [8,9]. It has been established that $T_c$ is related to the hole concentration in Cu-O sheets ($p_{sh}$) [10]. By adding equal amounts of CaO and CuO in non-superconducting, tetragonal $La_2Ba_2Cu_4O_z$ (x = 0.0), superconductivity is 'turned on' with maximum $T_c$ achieved for x = 0.5 (La-2125) compound.

In R-123 systems, magnetic moment of R does not contribute to superconducting properties, except for Ce, Tb and Pr [11]. Pr-substituted systems become non-superconductor above $x_{cr}$, where as, in the present study we observe increase in $T_c$ in with increase in Pr concentration.

2. **EXPERIMENTAL**

All the samples of LaPrCaBCO series (x = 0.1 – 0.5) were synthesized by solid state reaction method. The high purity (99.9+ %) starting compounds of $La_2O_3$, $BaCO_3$, $CaCO_3$, CuO and $Pr_6O_{11}$ were taken in stoichiometric quantities and grinded thoroughly using agate and mortar under acetone. The calcination and sintering of these compounds were carried out in pellet form in the temperature range of 930 – 940$^0$C for about 72 hrs with intermediate grindings. Thus prepared samples were annealed in oxygen at 500$^0$C for about 24 hrs and then slow cooled to room temperature.

The samples were characterized for their structural properties by X-ray diffraction (Cu-$K_\alpha$, $\lambda$ = 1.5405 Å) at room temperature. All the XRD patterns show single phase formation. The patterns were then fitted by Rietveld analysis using FULLPROF program [12]. The results show that there is no structural transition in the entire doping range, as all the samples are tetragonal, with decreasing unit cell parameters and volume. The transition temperature were recorded by four probe d. c. resistivity method and verified by d. c. susceptibility method, using a commercial SQUID magnetometer. Both the values are in good agreement, showing the good quality of samples. The oxygen content for all the samples were determined by iodometric double titration method. The hole concentration in the unit cell, and also in Cu-O sheets were calculated from oxygen content values using Tokura's method [13].

3. **RESULTS AND DISCUSSION**

The XRD Rietveld fitted patterns are shown in Figure 1. The refinement was done assuming an R-123 tetragonal unit cell and all the patterns fits well into this model. The occupancies were kept fixed, as per the site occupancy, and the fitting parameters like, scale factor, zero angle correction, atomic positions, half width parameters, unit cell parameters etc were varied to fit the observed and calculated profiles. The results of the analysis are listed in Table 1. Figure 2 shows the variation of unit cell parameters and volume with respect to increasing x. The decrease in unit cell parameters and cell volume can be attributed to the fact that, smaller ionic radii ions Pr and Ca are replacing La.

Since the starting model was assumed to be an R-123 tetragonal structure, the LaPrCaBCO (x = 0.1 – 0.5) were normalized to La-123. In this normalized form, La shares its site with both Pr and Ca; similarly Ba site is occupied by Ba, Ca and La. Increasing x, increases the percentage of Ca onto La and La onto Ba site. During initial refinements, Pr was kept full and fixed at La site. The refinement converged with good agreement between observed and calculated patterns.

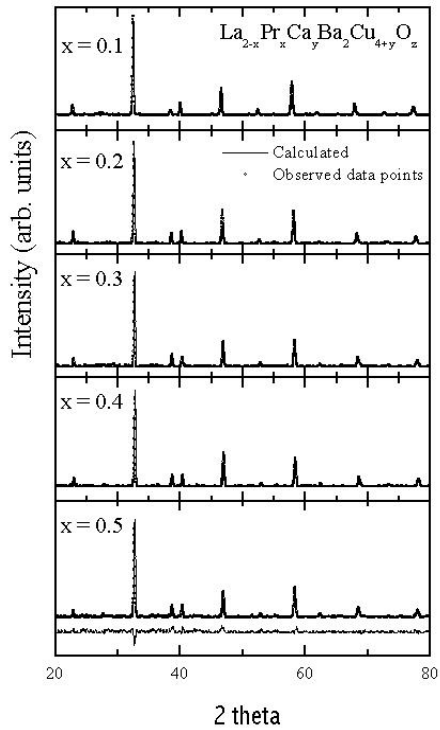

Fig. 1. *XRD Rietveld fitted pattern for all the samples. The open circles represent data points, and the line passing through them shows calculated pattern. The difference between observed and calculated pattern is shown for x = 0.5 sample*

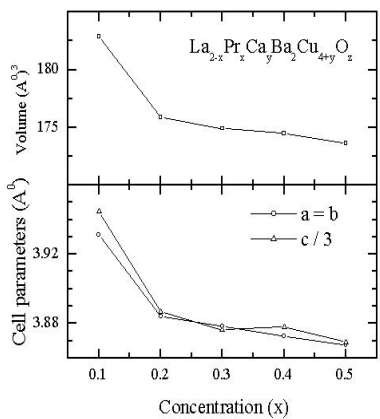

Fig.2. *Variation in unit cell parameters and volume with increasing Pr (x) content.*

Subsequent refinements with Pr distributed at both La and Ba site did not improve the fitting. Hence, Pr was kept full and fixed at La site during final analysis. There is also concomitant displacement of La onto Ba site. The Ca occupying La site and La onto Ba site can be viewed as hole doping and hole filling respectively. Also there is some portion of Ca at Ba site, which we assume may be counter balancing the hole filling by La. Looking at Table 1, out of total Ca in the sample, the Ca concentration increases from 35 % (for x = 0.1) to 68% (x = 0.5) at La site. The occupancy of Ca at La site increases from 39% - 41 %.

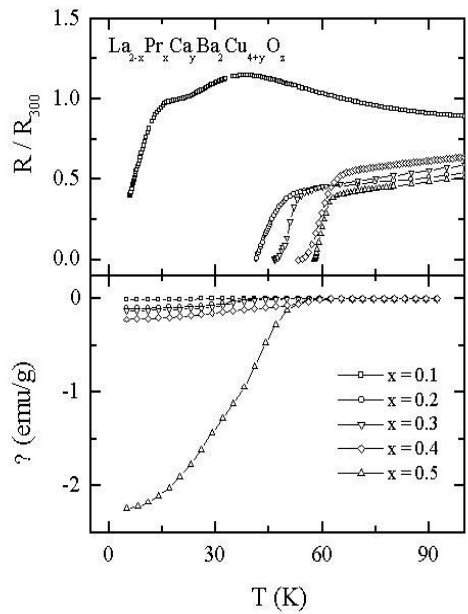

Fig. 3. *Transition temperatures for $La_{2-x}Pr_xCa_yBa_2Cu_{4+y}O_z$ compounds determined by resistivity and susceptibility as a function of temperature.*

Figure 3 shows the transition temperatures observed by electrical (resistance vs. temperature) and magnetic (susceptibility vs. temperature) methods. It is interesting to observe here that, with increasing Ca concentration at La site (as x increases), transition temperature increases, as can be seen in Figure 4.

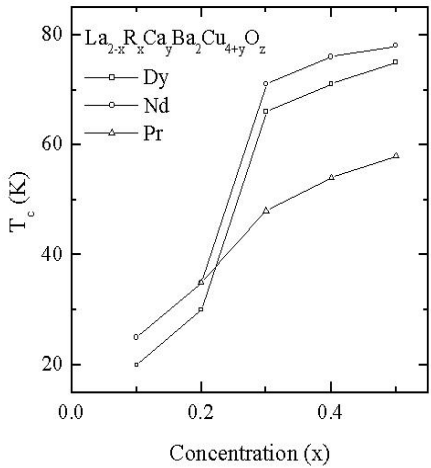

Fig. 4. *Variation in $T_c$ with increasing x, for R = Nd, Dy and Pr in $La_{2-x}R_xCa_yBa_2Cu_{4+y}O_z$*

The Figure 4 also shows, for the sake of comparison, the increase in $T_c$ for Nd and Dy substituted La-2125 compounds. The difference in maximum $T_c$ achieved by Pr and Nd & Dy systems is quite large (~ 20 K).

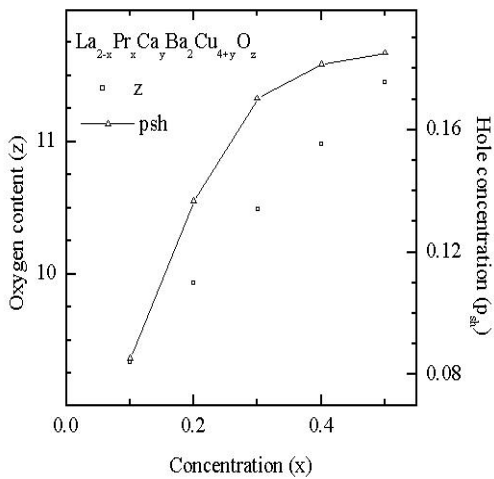

Fig. 5. *Change in oxygen content and hole concentration in sheets ($p_{sh}$).*

The Pr is known to hybridize with conducting $CuO_2$ plane and localize the mobile charge carriers. On the other hand, Ca is a known hole dopant in R-123 systems. In LaPrCaBCO system also, increasing x, increases oxygen content and hole concentration, as shown in Figure 5. Thus, we can feel that the rise in $T_c$ (but not up to maximum of La-2125 system i.e., ~ 78K) in LaPrCaBCO is due to the competition between hole filling and/or hybridization and hole doping which results from the simultaneous increase of both Pr and Ca with x.

**4. CONCLUSION**

The studies on LaPrCaBCO system, presents a unique case of increase in $T_c$ with the simultaneous increase of both Pr and Ca. The present study also highlights the role of Ca in contributing conducting holes to, otherwise non-superconductor, and 'turning on' the superconductivity. The present result of increase in $T_c$ with x is in consistence with our earlier observations, for different R, except for the maximum $T_c$ achieved. We feel that LaPrCaBCO compounds are possible candidates for further investigations in better understanding for the actual role Pr in high $T_c$ oxides.

**ACKNOWLEDGEMENTS**

The authors are thankful to Dr. D. C. Kundaliya and Prof. S. K. Malik for magnetic measurements. The work at Saurashtra University, Rajkot was carried out under the IUC-DAEF (India) project number CRS-M-88 of Dr. D. G. Kuberkar.

Table 1  Values obtained from the Rietveld analysis of $La_{2-x}Pr_xCa_yBa_2Cu_{4+y}O_z$ compounds for $x = 0.1 – 0.5$ ($y = 2x$)

| Parameters | Sample (x, y) | | | | |
|---|---|---|---|---|---|
| | (0.1, 0.2) | (0.2, 0.4) | (0.3, 0.6) | (0.4, 0.8) | (0.5, 1.0) |
| Space group | P4/MMM | P4/MMM | P4/MMM | P4/MMM | P4/MMM |
| a = b (Å) | 3.8988 (3) | 3.8824 (3) | 3.8724 (3) | 3.8650 (3) | 3.8712 (3) |
| c (Å) | 11.7235 (8) | 11.6698 (8) | 11.6417 (8) | 11.6336 (8) | 11.6388 (8) |
| La/Pr/Ca (½, ½, ½) | | | | | |
| $N_{La}$ | 0.821 | 0.630 | 0.451 | 0.476 | 0.274 |
| $N_{Pr}$ | 0.035 | 0.143 | 0.184 | 0.226 | 0.275 |
| $N_{Ca}$ | 0.035 | 0.094 | 0.279 | 0.226 | 0.395 |
| Ba (½, ½, z) | | | | | |
| z | 0.1804 | 0.1830 | 0.1819 | 0.1864 | 0.1868 |
| N | 1.433 | 1.328 | 1.242 | 1.255 | 1.120 |
| La @ Ba (N) | 0.505 | 0.509 | 0.594 | 0.505 | 0.560 |
| Ca @ Ba (N) | 0.076 | 0.090 | 0.142 | 0.255 | 0.180 |
| Cu (1) (0,0,0) | | | | | |
| N | 1.000 | 1.000 | 1.000 | 1.000 | 1.000 |
| Cu (2) (0,0,z) | | | | | |
| z | 0.3504 | 0.3521 | 0.3535 | 0.3536 | 0.3596 |
| N | 2.0000 | 2.0000 | 2.0000 | 2.0000 | 2.0000 |
| O (1) (0,½,0) | | | | | |
| N | 1.0562 | 1.3192 | 1.0100 | 0.9415 | 1.5050 |
| O (2) (0,0,z) | | | | | |
| z | 0.1527 | 0.1687 | 0.1594 | 0.1626 | 0.1575 |
| N | 2.0000 | 2.0000 | 2.0000 | 2.0000 | 1.3903 |
| O (4) (0,½,z) | | | | | |
| z | 0.3653 | 0.3799 | 0.3687 | 0.3622 | 0.3735 |
| N | 4.0000 | 4.0000 | 4.0000 | 4.0000 | 3.1823 |
| Total Oxygen | | | | | |
| (z' – in 123) | 7.0562 | 7.3192 | 7.0997 | 6.9415 | 7.0692 |
| (z – in 2125) | 9.8786 | 10.7348 | 10.8862 | 11.1064 | 11.7820 |
| Factors | | | | | |
| $\chi^2$ | 2.18 | 2.01 | 2.20 | 1.96 | 1.57 |
| $R_p$ | 22.8 | 28.9 | 30.6 | 35.2 | 32.7 |